\documentclass{ws-mpla}

\def\als{\alpha_s} 
\def\MS{\overline{\rm MS}}
\newcommand{\bea}{\begin{eqnarray}}
\newcommand{\eea}{\end{eqnarray}}
\newcommand{\nn}{\nonumber}

\begin{document}

\markboth{Xavier Garcia i Tormo}
{Determination of $\alpha_s$ from the QCD static energy}

\catchline{}{}{}{}{}

\title{Review on the determination of $\alpha_s$ from the QCD static energy}

\author{XAVIER GARCIA i TORMO}

\address{Albert Einstein Center for Fundamental Physics, Institut f\"ur Theoretische Physik, Universit\"at Bern,
  Sidlerstrasse 5, CH-3012 Bern, Switzerland\\
garcia@itp.unibe.ch}

\maketitle


\begin{abstract}
We review the determination of the strong coupling $\alpha_s$ from the
comparison of the perturbative expression for the Quantum
Chromodynamics static energy with lattice data. We collect here all
the perturbative expressions needed to evaluate the static energy at
the currently known accuracy.

\end{abstract}


\section{Introduction}
\begin{figure}
\centerline{\psfig{file=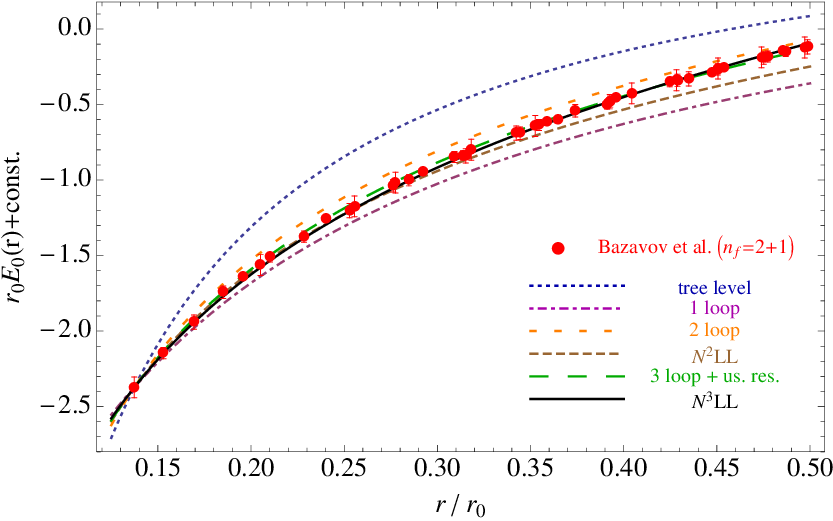,width=5.0in}}
\vspace*{8pt}
\caption{Comparison of the static energy calculated at different
  orders of accuracy (see Sec.~\ref{sec:pertexprE0} for explicit expressions) with lattice data for $n_f=2+1$
  \protect\cite{Bazavov:2011nk} ($n_f$ is the number of light
  flavors). The additive constant in the perturbative
  expression for the static energy is
  taken such that each curve coincides with the lattice data point at
  the shortest distance (see
  Eq.~(\ref{eq:Etilde})). $r_0\Lambda_{\MS}=0.70$ is used for all the curves ($\Lambda_{\MS}$ is the QCD scale, in the $\MS$ scheme,
  and $r_0$ is the lattice reference scale, see Sec.~\ref{sec:comp}). \protect\label{fig:E0nf3}}
\end{figure}

\begin{figure}
\centerline{\psfig{file=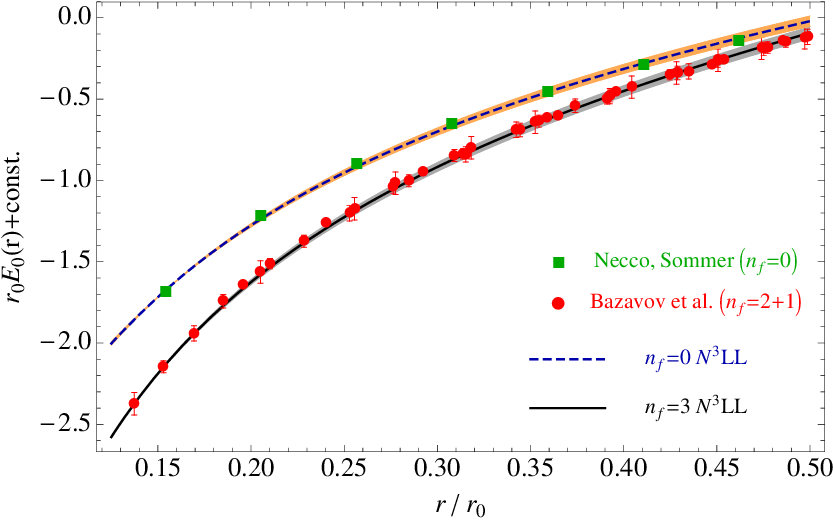,width=5.0in}}
\vspace*{8pt}
\caption{Comparison of the static energy with lattice data for $n_f=0$
  \protect\cite{Necco:2001xg} and $n_f=2+1$
  \protect\cite{Bazavov:2011nk} ($n_f$ is the number of light
  flavors). The theoretical curves include terms up to order
  $\alpha_s^{4+n}\ln^n\alpha_s$ (with $n\ge0$), what is referred to as
  N$^3$LL accuracy, see Sec.~\ref{sec:pertexprE0} for more details and
  explicit expressions. The additive constant in the perturbative
  expression for the static energy is
  taken such that each curve coincides with the corresponding lattice data point at
  the shortest distance (see Eq.~(\ref{eq:Etilde})). The bands are obtained by adding a term $\pm
  C_F\alpha_s^5/r$ to the N$^3$LL curves, and give an idea of the
  perturbative uncertainty of the results. $r_0\Lambda_{\MS}$=0.637 is
  used for the $n_f=0$ curve, and $r_0\Lambda_{\MS}=0.70$ for the
  $n_f=3$ one ($\Lambda_{\MS}$ is the QCD scale, in the $\MS$ scheme,
  and $r_0$ is the lattice reference scale, see Sec.~\ref{sec:comp}). \protect\label{fig:E0nf0nf3}}
\end{figure}

There has been much progress, in the last few years, in the perturbative
evaluation of the Quantum Chromodynamics (QCD) static energy $E_0(r)$,
i.e. the energy between a static quark and a static anti-quark
separated a distance $r$. Alongside, unquenched lattice computations of $E_0(r)$
at short distances have become available. These developments have made
manifest the ability of perturbative calculations in QCD to reproduce the
short-distance part of $E_0(r)$ calculated on the lattice, and have led to a determination of
the strong coupling $\alpha_s$ from the comparison of the
two~\cite{Bazavov:2012ka}, which is what we review here. As a
preface, let us illustrate the aforementioned progress by showing: (i)
a comparison of perturbative calculations for $E_0(r)$, at different
orders of accuracy, with short distance lattice data with three light
flavors~\cite{Bazavov:2011nk}, Fig.~\ref{fig:E0nf3}, and (ii)
short-distance lattice data for $E_0(r)$,
with zero~\cite{Necco:2001xg} and three~\cite{Bazavov:2011nk} light flavors, along with the corresponding
perturbative predictions at the highest accuracy known at present, Fig.~\ref{fig:E0nf0nf3}.
We can see from the figures that one can perfectly describe the short-distance
behavior of $E_0(r)$ obtained in the lattice, which can be considered
as an important landmark in our understanding of QCD. 

The rest of the paper is organized as follows: in
Sec.~\ref{sec:pertexprE0} we present the currently known terms in the
perturbative expansion of the static energy. Section \ref{sec:comp}
contains the comparison of lattice data with perturbation theory and
the corresponding extraction of $\alpha_s$. In Sec.~\ref{sec:concl} we conclude and
discuss the expected developments in the near future. The Appendix
collects color factors and beta function coefficients.

\section{Perturbative expression for the static energy}\label{sec:pertexprE0}
The present knowledge of $E_0(r)$ at short distances can be summarized
as follows
\begin{eqnarray}
E_0(r)&=&-\frac{C_F\als(1/r)}{r}\Bigg\{1+\frac{\als(1/r)}{4\pi}\tilde{a}_1
+\left(\frac{\als(1/r)}{4\pi}\right)^2
\tilde{a}_2
\nonumber\\
&&
+\left(\frac{\als(1/r)}{4\pi}\right)^3\left[a_3^L\log{\frac{C_A
    \als(1/r)}{2}} + \tilde{a}_3\right]
\nonumber\\
&&
+\left(\frac{\als(1/r)}{4\pi}\right)^4
\left[ a_{4}^{L2}\log^2 {\frac{C_A \als(1/r)}{2}}+a_{4}^{L}\log {\frac{C_A
  \als(1/r)}{2}}+ \tilde{a}_{4} \right] 
\nonumber\\
&&
+ \cdots \Bigg\}.
\label{eq:E0stat}
\end{eqnarray}
Non-analytic terms in $\alpha_s$ appear in $E_0(r)$, starting at order
$\als^4$, due to virtual emission of ultrasoft gluons~\cite{Appelquist:1977es}
(i.e. gluons with energy and momentum smaller  than $1/r$ that can
change the color state of the quark-antiquark pair  from singlet to
octet). 

Terms up to next-to-next-to-leading order (N$^2$LO), i.e. $\tilde{a}_1$
and $\tilde{a}_2$ in Eq.~({\ref{eq:E0stat}}), have been known for some
time \cite{Fischler:1977yf,Billoire:1979ih,Peter:1996ig,Peter:1997me,Schroder:1998vy,Kniehl:2001ju}. They read
\begin{equation}
\tilde{a}_1=:a_1+2\gamma_E\beta_0\quad;\quad\tilde{a}_2=:a_2 +\left(\frac{\pi^2}{3}+4\gamma_E^2\right)\beta_0^2
  +\gamma_E\left(4a_1\beta_0+2\beta_1\right),
\end{equation}
\bea
a_1&=&\frac{31}{9}C_A-\frac{20}{9}T_Fn_f,\\
a_2&=& \left({4343\over162}+4\pi^2-{\pi^4\over4}+{22\over3}\zeta(3)\right)C_A^2
-\left({1798\over81}+{56\over3}\zeta(3)\right)C_AT_Fn_f
\nn\\
&& -\left({55\over3}-16\zeta(3)\right)C_FT_Fn_f +\left({20\over9}T_Fn_f\right)^2.
\eea
More recently, the three-loop coefficient $\tilde{a}_3$ was computed
by two different groups~\cite{Smirnov:2008pn,Anzai:2009tm,Smirnov:2009fh}. It reads
\begin{eqnarray}
\tilde{a}_3 & =: & a_3+\Big(8\gamma_E^3+2\gamma_E\pi^2+16\zeta(3)\Big)
\beta_0^3+2\gamma_E\beta_2\nn\\
&&+\Big[\left(12\gamma_E^2+\pi^2\right)\beta_0^2+4\gamma_E\beta_1\Big]
a_1+\left[6a_2\gamma_E+\frac{5}{2}\left(4\gamma_E^2+\frac{\pi^2}{3}\right)\beta_1\right]\beta_0,
\end{eqnarray}
\begin{eqnarray}
a_3 & =: & a_3^{(3)}n_f^3+a_3^{(2)}n_f^2+a_3^{(1)}n_f+a_3^{(0)},\\
  a_3^{(3)} &=& - \left(\frac{20}{9}\right)^3 T_F^3
  \,,\\
  a_3^{(2)} &=&
  \left(\frac{12541}{243}
    + \frac{368}{3}\zeta(3)
    + \frac{64\pi^4}{135}
  \right) C_A T_F^2
  +
  \left(\frac{14002}{81}
    - \frac{416}{3}\zeta(3)
  \right) C_F T_F^2
  \,,\\
  a_3^{(1)} &=&
  \left(-709.717
  \right) C_A^2 T_F
 +
  \left(-\frac{71281}{162}
    + 264 \zeta(3)
    + 80 \zeta(5)
  \right) C_AC_F T_F
  \nn\\&&\mbox{}
  +
  \left(\frac{286}{9}
    + \frac{296}{3}\zeta(3)
    - 160\zeta(5)
  \right) C_F^2 T_F
 +
  \left(-56.83(1)
  \right) \frac{d_F^{abcd}d_F^{abcd}}{N_A}
  \, ,\\
a_3^{(0)} & = & 502.24(1) C_A^3 -136.39(12)
\frac{d_F^{abcd}d_A^{abcd}}{N_A}\nn\\
&&+\frac{8}{3}\pi^2C_A^3\left(-\frac{5}{3}+2\gamma_E+2\log2\right).\label{eq:a30}
\end{eqnarray}
Note that Refs.~\cite{Smirnov:2008pn,Anzai:2009tm,Smirnov:2009fh} use
a slightly different notation, in particular
Ref.~\cite{Smirnov:2009fh} uses $a_3^{(0)}$ to denote just the first
line of Eq.~(\ref{eq:a30}) above. The color
factors and beta function coefficients, which appear throughout the
paper, are collected in the Appendix, $\gamma_E=0.5772\dots$ is the
Euler constant, $\zeta(x)$ is the Riemann zeta function, and $n_f$ is
the number of light flavors. For convenience, we also give here the numerical values of the
$\tilde{a}_{1,2,3}$ coefficients for $N_c=3$
\bea
\hspace{-4mm}
\tilde{a}_1&=&23.032 - 1.8807 n_f, 
\\
\hspace{-4mm}
\tilde{a}_{2}&=& 1396.3 - 192.9 n_f + 4.9993 n_f^2,
\\
\hspace{-4mm}
\tilde{a}_{3}&=&
108654. - 21905.2 n_f + 1284.69 n_f^2 
- 20.6009 n_f^3.
\eea

The coefficients of the logarithmic terms in
Eq.~(\ref{eq:E0stat}) can be conveniently calculated within
the framework of the effective theory potential Non-Relativistic QCD
(pNRQCD)\cite{Pineda:1997bj,Brambilla:1999xf,Brambilla:2004jw}. They
read \cite{Brambilla:1999qa,Kniehl:1999ud,Pineda:2000gza,Brambilla:2006wp}
\begin{eqnarray}
a_3^L & = & \frac{16\pi^2}{3}C_A^3,\\
a_{4}^{L2} & = &
-\frac{16\pi^2}{3}C_A^3\beta_0,\label{eq:a4L2}\\
a_4^L &=&  16\pi^2C_A^3\left[a_1+2\gamma_E\beta_0 
+ T_F n_f \left( -\frac{40}{27} + \frac{8}{9} \log 2\right)
\right.
\nn\\
&&
\left.
+ C_A\left(\frac{149}{27}-\frac{22}{9}\log 2+\frac{4}{9}\pi^2\right)\right].
\label{eq:a4L}
\end{eqnarray}
pNRQCD can also be used to perform the resummation of these
logarithms. This was done done at leading order in
Ref.~\cite{Pineda:2000gza}, and at sub-leading order in
Ref.~\cite{Brambilla:2009bi}. When we include resummation of the
ultrasoft logarithms $E_0(r)$ reads\footnote{We thank Antonio Pineda for making us aware of a misprint in earlier versions of this formula~\cite{Ayala:2020odx}.}
\begin{eqnarray}
E_0(r)&=&-\frac{C_F\als(1/r)}{r}\Bigg\{1+\frac{\als(1/r)}{4\pi}\tilde{a}_1
+\left(\frac{\als(1/r)}{4\pi}\right)^2
\tilde{a}_2
\nonumber\\
&&
+\left(\frac{\als(1/r)}{4\pi}\right)^3\tilde{a}_3\Bigg\}\nn\\
&& +\frac{2}{3} C_F r^2
\left\{\frac{C_A}{2}\frac{\als(1/r)}{r}\left[1+\left(a_1+2\gamma_E\beta_0\right)\frac{\als(1/r)}{4\pi}\right]\right\}^3
\nn\\
&&\times\left( \frac{2}{\beta_0}  \ln \frac{\als(\mu)}{\als(1/r)} 
+ \left(\eta_0-\frac{1}{\pi}\left(-\frac{5}{6}+\log2\right)\right)\left[\als(\mu)- \als(1/r) \right] \right)\nn\\
&& -\frac{C_FC_A^3}{12\pi r}\als^3(1/r)\als(\mu)\log\frac{C_A\als(1/r)}{2 r\mu},
\label{eq:E0stat_usres}
\end{eqnarray}
with
\begin{equation}
\eta_0 = \frac{1}{\pi}\left(-\frac{\beta_1}{2\beta_0^2} + \frac{12B}{\beta_0}\right)\quad;\quad B = \frac{-10T_Fn_f + C_A(6\pi^2+47)}{108}.
\end{equation}
In Eq.~(\ref{eq:E0stat_usres}) we only display the terms
that are needed for next-to-next-to-next-to-leading logarithmic
(N$^3$LL) accuracy, where by N$^3$LL accuracy we mean that we include
terms up to order $\alpha_s^{4+n}\log^n\alpha_s$ ($n\ge0$). $\mu$ is
the ultrasoft factorization scale, which takes a natural value
$\mu\sim(C_A\als)/(2r)$; $E_0(r)$ is a physical observable and
therefore $\mu$ independent, i.e. the $\mu$ dependence in
Eq.~(\ref{eq:E0stat_usres}) cancels order by order. 

Let us recall at this point that in order to properly define the static limit of
QCD, or Heavy Quark Effective Theory, one needs to introduce a
residual mass term\cite{Beneke:1994sw}, whose typical size is
associated with the QCD hadronic scale, $\Lambda_{\rm QCD}$. This
residual mass term is inherited by pNRQCD. In the short-distance weak-coupling
regime we are considering here, it can be encoded in a matching
coefficient, that we denote by $\Lambda_s$, which should be added to
the expressions for $E_0(r)$ above. i.e. we have\cite{Brambilla:2009bi}
\begin{equation}
E_0(r)\to E_0(r)+\Lambda_s.
\end{equation}
The coefficient $\Lambda_s$ obeys ultrasoft renormalization group (RG)
equations in pNRQCD. The solution of the RG equations, at the order we will need it, reads
\begin{equation}\label{eq:lamsRG}
\Lambda_s(\mu)=K_1+K_2 \als^2(1/r)C_FC_A^2\frac{1}{\beta_0}\ln {\als(\mu)\over \als (1/r)},
\end{equation}
where $K_1$ and $K_2$ are dimension-one constants of order
$\Lambda_{\rm QCD}$. The term involving $K_2$ starts contributing at
N$^3$LL accuracy, since one counts $K_{1,2}\sim \Lambda_{\rm QCD}\sim\alpha_s^2/r\ll E_0\sim\alpha_s/r$ . When comparing the static energy with lattice data,
it is important to perform the comparison in a way that is not
affected by the presence of the so-called renormalon singularities~\cite{Aglietti:1995tg,Hoang:1998nz,Beneke:1998rk,Pineda:2002se}. We
will achieve this by explicitly working with a renormalon-free scheme
all the time. We choose the so-called renormalon subtracted (RS)
scheme introduced in Ref.~\cite{Pineda:2001zq}. In practice this means
that we need to include a subtraction term to the perturbative
expression for the static energy. If we compute the static energy at
$m$-loop order in perturbation theory, the subtraction term reads
\bea  
\textrm{RS\small{subtr.}} & = & R_{s} \, \rho \,  \sum_{n=1}^m \left(
\frac{\beta_0}{2\pi} \right)^n \als(\rho)^{n+1} \sum_{k=0}^2
d_{k} \frac{\Gamma(n+1+b-k)}{\Gamma(1+b-k)}\,,\label{eq:RSsubtr}
\eea
with
\bea
d_0 & = & 1\, ,\\
d_1 & = & \frac{\beta_1^2-\beta_2\beta_0}{4 b
   \beta_0^4}\, ,\\
d_2 & = & \frac{-2 \beta_0^4 \beta_3+4 \beta_0^3
   \beta_1 \beta_2+\beta_0^2
   \left(\beta_2^2-2 \beta_1^3\right)-2 \beta_0 \beta_1^2 \beta_2+\beta_1^4}{32 (b-1) b
   \beta_0^8}\, ,\\
b & = & \frac{\beta_1}{2\beta_0^2}\, .
\eea
$R_s$ in Eq.~(\ref{eq:RSsubtr}) is the normalization of the first
renormalon singularity, it can be computed approximately using the
procedure of Ref.~\cite{Lee:1999ws}; $\rho$ is a dimensional scale
with a natural value around the center of the range of distances we consider, the
presence of a dimensional scale is inherent in all schemes that
explicitly cancel renormalon singularities. When we consider N$^3$LL
accuracy, a corresponding subtraction term is also needed for the term
in curly braces in the third line of Eq.~(\ref{eq:E0stat_usres}) (this
term arises from a difference of the color octet and color singlet
potentials, therefore the renormalon subtraction here also involves an
octet normalization constant $R_o$). For the
renormalon to cancel order by order in $\alpha_s$, one needs to expand
$\alpha_s(\rho)$ in terms of $\alpha_s(1/r)$ or vice versa, in order
to have a single expansion parameter in the final expression for the
static energy. Here we choose to expand $\alpha_s(1/r)$ in terms of
$\alpha_s(\rho)$, note that in this case the explicit numerical value
of the renormalon normalization constants is irrelevant for the lattice comparison in the next section.

The final expression for the static energy that we need to use is therefore
given by
\begin{equation}\label{eq:E0}
E_0(r)=\left[\textrm{Eq.~}(\ref{eq:E0stat_usres})\right]-\textrm{RS\small{subtr.}}+\Lambda_s,
\end{equation}
where it is understood that each of the three terms in
Eq.~(\ref{eq:E0}) is taken at the order needed to obtain the desired
accuracy. Note that, in order to simplify the notation and to avoid a
proliferation of symbols in the paper, earlier we denoted
Eq.~(\ref{eq:E0stat_usres}) (and Eq.~(\ref{eq:E0stat})) also by
$E_0(r)$, i.e. we use $E_0(r)$ as a generic denotation for the static
energy, without specifying in the notation if ultrasoft resummation is
performed or not, whether the perturbative expansion incorporates an
explicit renormalon subtraction, or the presence of a residual mass
term (which would only be absent in a purely perturbative result in $\MS$-like schemes).

\section{Comparison with lattice data and extraction of $\alpha_s$}\label{sec:comp}
We can now compare the perturbative expressions for the static energy
in the previous section with lattice data with three light
flavors. This comparison allows us to extract the value of the QCD
scale $\Lambda_{\MS}$ (in the $\MS$ scheme), upon which the
perturbative expressions depend. In order to obtain this
extraction, we assume that perturbation theory, after implementing a cancellation of the leading
renormalon singularity, is enough to describe lattice data in the range of
distances we study. 

We employ the $n_f=2+1$ lattice data for the
static energy obtained in Ref.~\cite{Bazavov:2011nk}. This lattice
computation used a combination of tree-level improved gauge
action and highly-improved staggered quark action
\cite{Follana:2006rc}. It employed the physical value for the strange-quark mass $m_s$
and light quark masses equal to $m_s/20$, which correspond to a pion
mass of about 160 MeV in the continuum limit, very close to the
physical value. The computation was performed for a wide range of gauge
couplings, and was corrected for lattice artifacts. At each value of the gauge coupling one calculates the scale parameters $r_0$ and $r_1$ defined
in terms of the static energy $E_0(r)$ as follows \cite{Sommer:1993ce,Aubin:2004wf}
\begin{equation}
r^2 \frac{d E_0(r)}{d r}|_{r=r_0}=1.65,~~~r^2 \frac{d E_0(r)}{d r}|_{r=r_1}=1.
\end{equation}
The values of $r_0$ and $r_1$ were given in Ref. \cite{Bazavov:2011nk}
for each gauge coupling. The static energy can be calculated in units
of $r_0$ or $r_1$. For the present analysis, we only use lattice data
for $r<0.5r_0$, where perturbation theory should be reliable. Since we have lattice data points down to $r=0.14r_0$, this means that we
are studying the static energy in the 0.065~fm$\lesssim r
\lesssim$0.234~fm distance range, in physical units. The static
energy has an additive ultraviolet renormalization (the self energy of
the static sources) and one needs to normalize the results 
calculated at different lattice spacings to a common value at a
certain distance (as an alternative to that one can also take a
derivative and compute the force). The static energy in units of $r_0$ is fixed to 0.954 at
$r = r_0$. For additional details about the
lattice data see Refs.~\cite{Bazavov:2011nk,Bazavov:2012ka}. The
adequate quantity to plot in order to compare with lattice data is:
\begin{equation}
\label{eq:Etilde}
E_0(r)-E_0(r_{\rm min})+E_0^{\rm latt.}(r_{\rm min})=E_0(r)+\textrm{const.},
\end{equation}
where $r_{\rm min}$ is the shortest distance at which lattice data is
available, and $E_0^{\rm latt.}(r_{\rm min})$ is the value of the
lattice data at that distance. Note that then, by construction, all
perturbative curves coincide with the lattice point at the shortest
distance available. Therefore, for instance, the N$^3$LL curves that are shown in
Fig.~\ref{fig:E0nf0nf3} are given by Eq.~(\ref{eq:Etilde}) with $E_0(r)$ from Eq.~(\ref{eq:E0}), taking each of the three terms in
that last equation at N$^3$LL accuracy (with $K_2$ fitted to the
lattice data), recall also that we always express everything as an
expansion in terms of $\alpha_s(\rho)$. Corresponding expressions hold
for the rest of the curves. Let us also mention that in principle one can include finite
strange-quark mass effects at one loop~\cite{Hoang:2000fm,Eiras:2000rh} in Eq.~(\ref{eq:E0stat_usres})
or Eq.~(\ref{eq:E0stat}), but they turn out to be negligible.

We can now search for the values of
$\Lambda_{\MS}$ that are allowed by lattice data. The guiding
principle we follow to achieve this is that the agreement with lattice should improve when the
perturbative order of the calculation is increased. A procedure to
perform the extraction following these guidelines was devised in
Ref.~\cite{Brambilla:2010pp}, where it was applied to extract
$r_0\Lambda_{\MS}$ for the $n_f=0$ case. It consists of the steps
described next. 

First, to obtain the central value for $r_0\Lambda_{\MS}$ we:
\begin{enumerate}
\item Let $\rho$ vary by $\pm 25\%$ around its natural value at the
  center of the range where we have lattice data.
\item For each value of $\rho$, and at each order in the perturbative
  expansion of the static energy, we perform a fit to the lattice
  data ($r_0\Lambda_{\MS}$ is the parameter of each of the fits).
\item We select the $\rho$ values for which the reduced $\chi^2$ of
  the fits decreases when increasing the number of loops of the perturbative calculation.
\end{enumerate}
Then we consider the set of $r_0\Lambda_{\MS}$ values in the $\rho$
range we have obtained and take their average, using the inverse
reduced $\chi^2$ of each fit as weight. This gives the central value
for $r_0\Lambda_{\MS}$. We can do that at different orders of
accuracy, and obtain the results shown in Tab.~\ref{tab:Lam}.
\begin{table}[h]
\tbl{Values of $r_0\Lambda_{\MS}$ obtained at different levels of
  accuracy. ``N$^2$LL'' stands for next-to-next-to leading-logarithmic
  and ``3
  loop + us. res.''  stands for three loop plus leading ultrasoft logarithmic resummation.}
{\begin{tabular}{ccccc} \toprule
\hspace{1cm} & Accuracy & \hspace{4cm} & $r_0\Lambda_{\MS}$ & \hspace{1cm} \\
\hline &tree level& & $0.395$ &\\
&1 loop && $0.848$ &\\
&2 loop && $0.636$ &\\
&N$^2$LL && $0.756$ &\\
&3 loop && $0.690$ &\\
&3 loop + us. res.& & $0.702$&\\ \botrule
\end{tabular}\label{tab:Lam} }
\end{table}
Note that the last row of the column is at three loop plus leading
logarithmic resummation accuracy. If we perform the fits at
N$^3$LL accuracy, then an additional constant, $K_2$ in Eq.~(\ref{eq:lamsRG})
above, enters in them and also needs to be fitted. If we try to do
that we find that, with the present
lattice data, the $\chi^2$ as a
function of $r_0\Lambda_{\MS}$  is very flat; which means that at
present we
cannot improve our extraction of $r_0\Lambda_{\MS}$ by including the
fits at N$^3$LL accuracy in the analysis. Consequently, we take the
numbers in the last row of Tab.~\ref{tab:Lam} as our best result.

Then, to associate an error to this number we do the following. On the
basis that the error associated to the result should reflect the
uncertainties from unknown higher perturbative orders, we consider the
weighted standard deviation in the range of $\rho$ obtained above, and
the difference with the weighted average computed at the previous
perturbative order. We take those two numbers as errors of our result,
and add them linearly. We obtain $r_0\Lambda_{\MS}=0.7024\pm0.0011\pm0.0665=0.70\pm0.07$,
where the first error is due to the weighted standard deviation, and the
second to the difference with the two-loop result. Note that assigning
the difference with the result at the previous order as an error is a
quite conservative estimate. To further assess possible systematic errors stemming from our
procedure, we have redone the analysis using $p$-value weights and
using constant weights. We find similar results, and in the final
result quote an error that covers the whole range spanned by the three
analyses. A partial additional cross-check of the result can be
performed by redoing the whole analysis with the static energy normalized in units
of the scale $r_1$, rather than $r_0$. Note that this is a
cross-check, and not just a trivial re-scaling, because the
systematics and errors entering the lattice analysis normalized in
units of $r_0$ or $r_1$ are different. When we do that, we do find
consistent results.

From the above discussion, our final result for $r_0\Lambda_{\MS}$ reads
\begin{equation}\label{eq:Lambda}
r_0\Lambda_{\MS}=0.70\pm0.07,
\end{equation}
which using $r_0=0.468\pm0.004$ fm \cite{Bazavov:2011nk} corresponds to
\begin{equation}\label{eq:as1p5}
\alpha_s\left(\rho=1.5{\rm GeV},n_f=3\right)=0.326\pm0.019,
\end{equation}
the uncertainty in $r_0$ is negligible in the final error
above. $\rho\sim1.5$~GeV, which corresponds to the center of the range where we have
lattice data, is the natural scale of our $\alpha_s$ determination. When we evolve this value to the scale of the $Z$ mass, $M_Z$, we obtain
\begin{equation}\label{eq:asMZ}
\alpha_s\left(M_Z,n_f=5\right)=0.1156^{+0.0021}_{-0.0022},
\end{equation}
where we have used the \verb|Mathematica| package \verb|RunDec| \cite{Chetyrkin:2000yt} to obtain
the above number (4 loop running, with the charm-quark mass equal to
1.6 GeV and the bottom-quark mass equal to 4.7 GeV).

Before concluding, let us mention that some studies for the $n_f=2$ case have been
presented in Refs.~\cite{Jansen:2011vv,Leder:2011pz}; and that
previous analyses for $n_f=0$, using perturbative expressions for the
static energy at the two-loop level, include Refs.~\cite{Sumino:2005cq,Necco:2001gh}.

\section{Conclusions and outlook}\label{sec:concl}
\begin{figure}
\centerline{\psfig{file=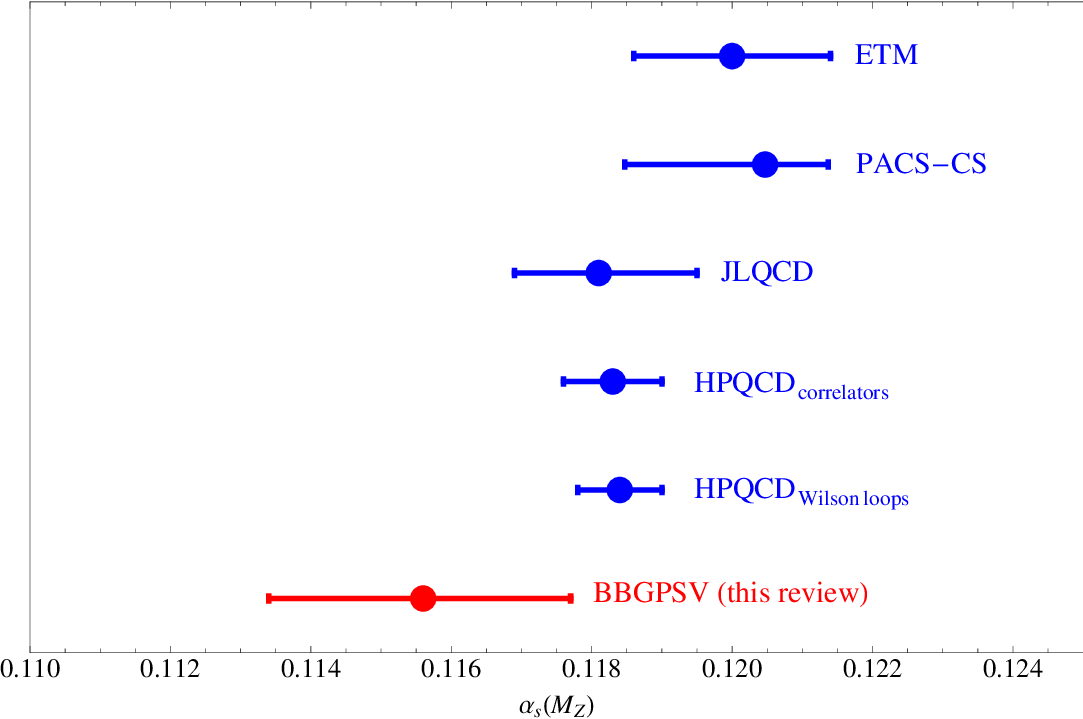,width=5.0in}}
\vspace*{8pt}
\caption{Comparison of the result for $\alpha_s(M_Z)$ in Eq.~(\protect\ref{eq:asMZ_2}) with other recent lattice determinations. The
  references are: HPQCD \protect\cite{McNeile:2010ji}, JLQCD
  \protect\cite{Shintani:2010ph}, PACS-CS \protect\cite{Aoki:2009tf}, ETM \protect\cite{Blossier:2012ef}.}\protect\label{fig:comp_als_latt}
\end{figure}

\begin{figure}[p]
\centerline{\psfig{file=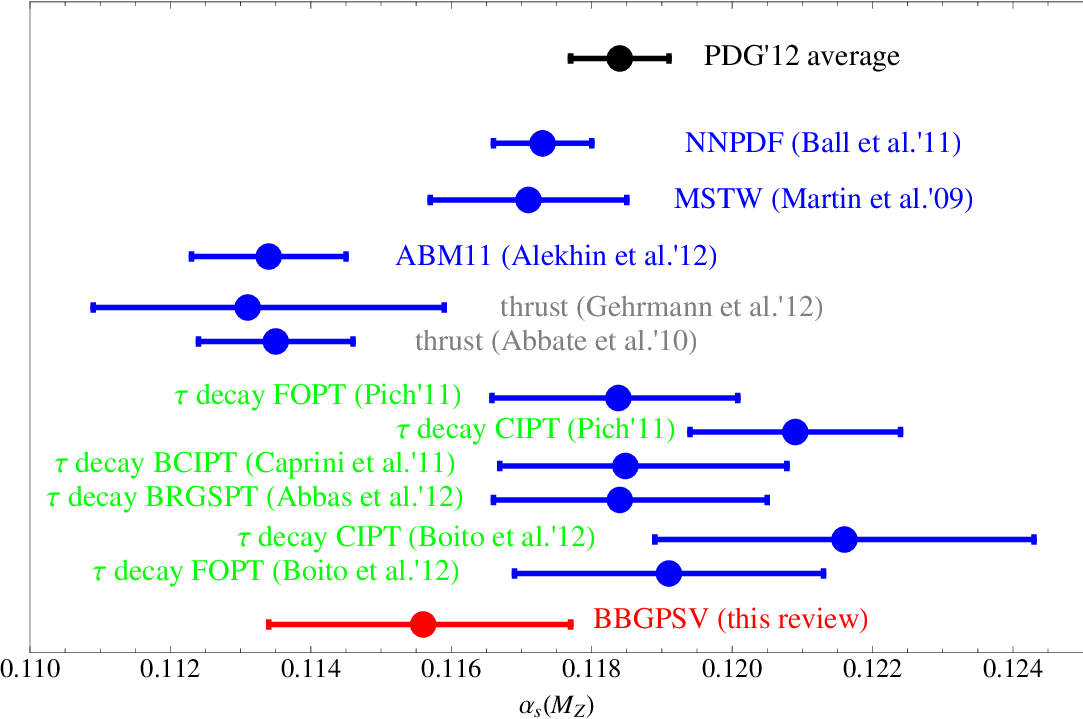,width=5.0in}}
\vspace*{8pt}
\caption{Comparison of the result for $\alpha_s(M_Z)$ in
  Eq.~(\protect\ref{eq:asMZ_2}) with a few other recent non-lattice $\alpha_s$
  determinations. We include results from $\tau$ decays (Boito {\it et
    al.}~\protect\cite{Boito:2012cr}; Abbas {\it et
    al.}~\protect\cite{Abbas:2012fi,Abbas:2012py}; Caprini {\it et al.}~\protect\cite{Caprini:2011ya};
  Pich~\protect\cite{Pich:2013sqa}), thrust (Abbate {\it et
    al.}~\protect\cite{Abbate:2010xh}; Gehrmann {\it et
    al.}~\protect\cite{Gehrmann:2012sc}), and parton distribution
  function (PDF) fits (ABM11~\protect\cite{Alekhin:2012ig},
  MSTW~\protect\cite{Martin:2009bu}, NNPDF~\protect\cite{Ball:2011us};
  note that in this case the error bars do not include effects from
  unknown higher-order perturbative corrections), along with the
  PDG average~\protect\cite{Beringer:1900zz}.}\protect\label{fig:comp_als_altr}
\end{figure}

We have reviewed the determination of $\alpha_s$ from the comparison
of lattice data with perturbative expressions for the QCD static
energy. This determination was possible due to the recent advances in
both the perturbative computation and the lattice evaluation of the
static energy. It can be viewed as a nice example where a three loop
computation is needed, and leads to an improved determination of a
Standard Model parameter, as was expected to happen (see for instance Ref.~\cite{Brambilla:2010cs}). We have collected here (in Sec.~\ref{sec:pertexprE0})
all the perturbative expressions needed to evaluate the static energy
at the currently known accuracy, which were scattered over different papers. The final result of the current analysis is 
\begin{equation}\label{eq:asMZ_2}
\alpha_s\left(M_Z,n_f=5\right)=0.1156^{+0.0021}_{-0.0022},
\end{equation}
which uses lattice data in the 0.8-2.9~GeV energy range. The result is mostly compatible with other recent lattice determinations of
$\alpha_s$, although the central value is a bit lower, see
Fig.~\ref{fig:comp_als_latt} for a graphical comparison with recent
lattice results. 
In Fig.~\ref{fig:comp_als_altr} we illustrate where the result in
Eq.~(\ref{eq:asMZ_2}) lays with respect to a few other recent
non-lattice determinations. 
New lattice
data for the static energy, also at shorter distances, will be available in
the near future. Therefore, an updated result for $\alpha_s$, with, in
principle, reduced errors, can be expected to appear in the next few months.

\appendix

\section{Color factors and beta function coefficients}\label{sec:app}
The color factors that appear in the paper read
\[
C_F=T_F\frac{N_c^2-1}{N_c}\quad;\quad C_A=N_c\quad;\quad T_F=\frac{1}{2}\quad;
\]
\[
\frac{d_F^{abcd}d_A^{abcd}}{N_A}=\frac{N_c^3+6N_c}{48}\quad ; \quad \frac{d_F^{abcd}d_F^{abcd}}{N_A}=\frac{18-6N_c^2+N_c^4}{96N_c^2}\quad;
\]
\begin{equation}
\frac{d_A^{abcd}d_A^{abcd}}{N_A}=\frac{N_c^4+36N_c^2}{24},
\end{equation}
where $N_c$ is the number of colors.

We define the beta function as
\begin{equation}
\als\beta(\als) = \frac{d\,\als(\nu)}{d\ln \nu} = -\frac{\als^2}{2\pi}
\sum_{n=0}^\infty \left( \frac{\als}{4\pi} \right)^n \beta_n=-2\alpha_s\left[\beta_0\frac{\alpha_s}{4\pi}+\beta_1\left(\frac{\alpha_s}{4\pi}\right)^2+\cdots\right],
\end{equation}
where\cite{vanRitbergen:1997va}
\begin{eqnarray}
\beta_0 & = & \frac{11}{3}C_A-\frac{4}{3}T_Fn_f,\\
\beta_1 & = & \frac{34}{3}C_A^2-\frac{20}{3} C_A n_f T_F-4 C_F n_f T_F,\\
\beta_2 & = & \frac{2857}{54}C_A^3+
   \left(-\frac{1415}{27}C_A^2-\frac{205}{9}C_A C_F+2C_F^2\right) n_f
   T_F\nn\\
&&+\left(\frac{158}{27}C_A+\frac{44}{9}C_F\right)n_f^2 T_F^2,\\
\beta_3 & = & \left(\frac{150653}{486}-\frac{44}{9}\zeta (3)\right)C_A^4+\left(\frac{136}{3}\zeta (3)-\frac{39143}{81}\right) C_A^3 n_f
   T_F\nn\\
&&+\left(\frac{7073}{243}-\frac{656}{9}\zeta (3)\right) C_A^2 C_F
   n_f T_F\nn\\
&& +\left(\frac{352}{9}\zeta (3)-\frac{4204}{27}\right) C_A C_F^2 n_f
   T_F\nn\\
&&+46 C_F^3 n_f T_F +\left(\frac{224}{9}\zeta (3)+\frac{7930}{81}\right) C_A^2 n_f^2
   T_F^2\nn\\
&&+\left(\frac{448}{9}\zeta (3)+\frac{17152}{243}\right) C_A C_F n_f^2T_F^2\nn\\
&&+\left(\frac{1352}{27}-\frac{704}{9}\zeta (3)\right) C_F^2 n_f^2
   T_F^2 +\frac{424}{243} C_A n_f^3
   T_F^3+\frac{1232}{243} C_F n_f^3 T_F^3\nn\\
&&+\left(\frac{512}{9}-\frac{1664}{3}\zeta (3)\right) n_f
   \frac{d_F^{abcd} d_A^{abcd}}{N_A}\nn\\
&&+\left(\frac{512}{3}\zeta
   (3)-\frac{704}{9}\right) n_f^2 \frac{
 d_F^{abcd}d_F^{abcd}}{N_A}+\left(\frac{704}{3}\zeta
   (3)-\frac{80}{9}\right) \frac{d_A^{abcd}d_A^{abcd}}{N_A}.
\end{eqnarray}

\section*{Acknowledgments}
It is a pleasure to thank Alexei Bazavov, Nora Brambilla, P\'eter
Petreczky, Joan Soto, and Antonio Vairo for collaboration on the work
reported in this review. This work is supported by the Swiss National Science Foundation (SNF)
under the Sinergia grant number
CRSII2\underline{ }141847\underline{ }1.

\end{document}